\documentclass[aip,pop,twocolumn,reprint,numerical,superscriptaddress,nofootinbib]{revtex4-1}
\usepackage{amssymb, amsmath,framed,amsthm}

\makeatletter
\@ifpackageloaded{stix} % if 'stix' packages is loaded
{
	\newcommand{\bm}{\boldsymbol}
}{                      % if not loaded
	\renewcommand{\bm}{\boldsymbol}
	\usepackage{mathrsfs}
	\usepackage{txfonts}              % Times-based fonts for text and math (equations become tight)
}
\makeatother

\usepackage[unicode=true,pdfusetitle,bookmarks=true,bookmarksnumbered=false,bookmarksopen=false,breaklinks=false,pdfborder={0 0 0},backref=false,colorlinks=true]{hyperref}
\hypersetup{citecolor=blue,filecolor=blue,linkcolor=blue,urlcolor=blue}
\usepackage{color,soul}
\usepackage{siunitx} % Comprehensive (SI) units
\usepackage{graphicx}% Include figure files
\usepackage{dcolumn}% Align table columns on decimal point
\usepackage{cancel}

\newcommand{\p}{\partial}
\newcommand{\f}[2]{\frac{#1}{#2}}
\newcommand{\mr}[1]{\mathrm{#1}}

\newcommand{\ds}{\displaystyle}
\newcommand{\dd}[2]{\frac{\rmd#1}{\rmd#2}}
\newcommand{\pp}[2]{\frac{\p #1}{\p #2}}

\newcommand{\nbl}{\nabla}
\makeatletter
\newcommand{\vast}{\bBigg@{4}}
\newcommand{\Vast}{\bBigg@{5}}
\makeatother

\DeclareMathAlphabet{\mathpzc}{OT1}{pzc}{m}{it}
\newcommand{\rmb}{\mathrm{b}}
\newcommand{\rmc}{\mathrm{c}}
\newcommand{\rmd}{\mathrm{d}}

\newcommand{\rmm}{\mathrm{m}}

\begin{document}

\title{Inward Diffusion and Acceleration of Particles Driven by Turbulent Fluctuations in Magnetosphere}

\author{Y. Ushida}
\affiliation{Faculty of Engineering, The University of Tokyo, Hongo, Tokyo 113-0033, Japan}
\author{Y. Kawazura}
\affiliation{Graduate School of Frontier Sciences, The University of Tokyo, Kashiwa, Chiba 277-8561, Japan}
\author{N. Sato}
\affiliation{Graduate School of Frontier Sciences, The University of Tokyo, Kashiwa, Chiba 277-8561, Japan}
\author{Z. Yoshida}
\affiliation{Graduate School of Frontier Sciences, The University of Tokyo, Kashiwa, Chiba 277-8561, Japan}

\date{\today}

% \pacs{95.30.Qd, 95.30.Sf, 52.27.Ny, 03.30.+p, 03.50.-z}% PACS, the Physics and Astronomy Classification Scheme.
% 95. Fundamental astronomy and astrophysics; instrumentation, techniques, and astronomical observations
	% 95.30.Qd	Magnetohydrodynamics and plasmas
	% 95.30.Sf	Relativity and gravitation

% 03. Quantum mechanics, field theories, and special relativity
	% 03.30.+p	Special relativity
	% 03.50.-z	Classical field theories

% 47. Fluid dynamics
	% 47.10.A-	Mathematical formulations
	% 47.75.+f	Relativistic fluid dynamics

% 52. Physics of plasmas and electric discharges
	% 52.27.Ny	Relativistic plasmas
	% 52.30.Cv	Magnetohydrodynamics

\begin{abstract}
	Charged particles in a magnetosphere are spontaneously attracted to a planet while increasing their kinetic energy via inward diffusion process.
	A constraint on particles' micro-scale adiabatic invariants restricts the class of motions available to the system, giving rise to a proper frame on which particle diffusion occurs. 
	We investigate the inward diffusion process by numerical simulation of particles on constrained phase space.
	The results reveal the emergence of inhomogeneous density gradient and anisotropic heating, which is consistent with spacecraft observations, experimental observations, and the recently formulated diffusion model on the constrained phase space.
\end{abstract}

\maketitle

%-------------------------------------------------------
% INTRODUCTION
%-------------------------------------------------------
% \section{INTRODUCTION}%
Magnetospheres are the prototypical systems that demonstrate spontaneous confinement of plasmas by magnetic force. 
Since magnetic force is free of mechanical work, its effect does not appear as an energy term in the Boltzmann distribution (which is in marked contrast with the gravitational confinement created by a star).
Instead, the magnetic field manifests itself as topological constraints in the dynamics and equilibrium structures; 
for example, see Ref.~\onlinecite{Yoshida2014} for a recent formulation of magnetic confinement in the perspective of phase-space foliation.

The self-organization of a magnetospheric plasma confinement, both in astronomical magnetic dipoles~\cite{selesnick1987voyager,persoon2013plasma} and laboratory ones~\cite{yoshida2010magnetospheric,boxer2010turbulent,saitoh2011high}, requires a spontaneous mechanism that `creates' density gradients.  
As a concomitant effect, particles are accelerated (heated) as they climb up the density gradients~\cite{dessler2002physics}; 
conservation of first and second adiabatic invariants along the inward displacement increases particle's kinetic energy.
The Van Allen radiation belt is believed to be the product of such process~\cite{kellogg1959van,brice1973jupiter,coroniti1974energetic}
(electrons in an ultra-relativistic regime are created by non-local acceleration mechanisms, such as wave particle interaction~\cite{Chen,Horne,thorne2013rapid}).
Recently, the inward diffusion heating was observed in laboratory magnetosphere experiments~\cite{Kawazura2015,Kawazura2016}.
As mentioned above, magnetic field does not produce a potential energy (unlike gravity or electrostatic force); 
hence the concentration and acceleration are not due to centripetal force.  
The driving force for such `up-hill diffusion'~\cite{schulz2012particle} and acceleration may come from some fluctuations. 
The key element of the mechanism is, then, the symmetry breaking that selects the preferential direction for particles to penetrate. 
The symmetry breaking appears in the metric of the phase space; 
the root cause of an inhomogeneous metric is the topological constraint imposed on magnetized particles by the adiabatic invariants such as magnetic moments~\cite{Gubar1989,Yoshida2014,Hasegawa2005}. 

The early theoretical studies developed an empirical Fokker--Planck type diffusion model on a phase space spanned by adiabatic invariants~\cite{birmingham1967charged} and explained planetary radiation belt with inhomogeneous density gradients (see Ref.~\onlinecite{walt2005introduction} and references therein).
This theory was later developed into a unified model according to which the number of particles contained in each magnetic flux tube tends to be homogenized, with the result that a peaked density profile is formed inwardly where the flux tube volume diminishes~\cite{Hasegawa2005}.
% This model is applicable for a tokamak configuration~\cite{baker1998density,baker2002use}. 
% However, since tokamak magnetic field is more homogeneous than dipole field, the density profile in a tokamak becomes also rather homogeneous~\cite{kesner2011fluctuation}. 
% The same Fokker--Planck type equation showed that a temperature profile becomes inhomogeneous, and its profile is predicted by equalized entropy density per unit magnetic flux~\cite{kesner2011fluctuation}.
Although this model is applicable to a tokamak configuration~\cite{baker1998density,baker2002use}, the density profile in a tokamak is homogeneous compared to that of a dipole field, since the magnetic field of the former is more homogeneous~\cite{kesner2011fluctuation}. 
The aforementioned Fokker--Plank type equation also revealed that an inhomogeneous temperature profile is obtained by equalising entropy density per unit magnetic flux~\cite{kesner2011fluctuation}.
The results in the literature are consistent with a gyrokinetic simulation in a dipole configuration~\cite{kobayashi2010particle}.
Recently, the empirical kinetic equation was reformulated in more rigorous manner~\cite{Sato2015} based on the idea of a phase space foliation~\cite{Yoshida2014}.  
The numerical simulations of the model revealed the inhomogeneous density profile~\cite{Sato2015} and anisotropic heating~\cite{sato2015self,sato2016up}.

The aim of this work is to put the inward diffusion process into the test by particle simulation.  
By doing so, we elude a stochastic modeling and examine the inward diffusion process not in terms of diffusion coefficient, but amplitude and time scale of perturbation.
We also examine the marginal regimes of adiabatic invariances of both cyclotron and bounce motions.  
The latter is of special interest, because 
the conservation of the second adiabatic invariant may increase the magnetic-field aligned temperature as the particles diffuse inward (Fermi acceleration), if the bounce action is conserved.  
The results are compared with the diffusion model (Fokker-Planck equation) based on the same ansatz of topological constraint~\cite{Sato2015}.

%-------------------------------------------------------
% HAMILTONIAN FORM AND PHASE-SPACE FOLIATION
%-------------------------------------------------------
% \section{HAMILTONIAN FORM AND PHASE-SPACE FOLIATION}%
We construct the model of magnetospheric particle motion upon the magnetic coordinate $(\ell, \psi, \theta)$, which may be interpreted as the proper frame on which particle diffusion occurs~\cite{Sato2015}. 
Here, an axisymmetric magnetic field with no toroidal component may be written as $\bm{B} = \nbl\psi\times\nbl\theta$, allowing the denotation of the coordinate by $\ell$ along the magnetic field lines, the magnetic flux function $\psi$, and the toroidal angle $\theta$. 
In a strong enough magnetic field, the canonical angular momentum  $P_\theta = mrv_\theta + q\psi$ is dominated by $q\psi$, constraining the radial position of particles to the magnetic surface defined by $\psi$, where $m$ and $q$ are the particle mass and charge, $v_\theta$ is the toroidal drift velocity and $r$ is the radial coordinate. 
$P_\theta$ is the adiabatic invariant corresponding to drift motion which is the most macroscopic among the three characteristic periodic motions in magnetospheres (i.e. the cyclotron, bounce, and drift motion). 
% The relative vulnerability to perturbation of drift motion to the other periodic motions points to the interpretation of the magnetic coordinate as the proper frame on which particle diffusion occurs~\cite{Yoshida2014}.

The macroscopic motion of a particle may be derived as the particle motion on a foliation of phase space, where the class of motions available to the system is restricted. 
This is done by applying general Hamiltonian mechanics~\cite{Morrison1998}. 
Macroscopic particle motion is described as a non-canonical Hamiltonian system with a foliated phase space by modifying the Poisson operator and separating micro-scale variables~\cite{Yoshida2014}. 
In the case of magnetospheres, choosing the adiabatic invariants of cyclotron, bounce, and drift motion and their angle variables as the canonical phase space $\bm{z} = (\mu,\,\theta_\rmc;\,J_{||},\,\theta_\rmb;\,P_\theta,\,\theta)$ and writing the Hamiltonian as $H = \mu\omega_\rmc + J_{||}\omega_\rmb + q\phi$, the cyclotron motion variables may be separated.
Here, the magnetic moment $\mu$, the bounce action $J_{||}$ and $P_\theta$ are the adiabatic invariants of the cyclotron, bounce, and drift motion, with $\theta_\rmc$, $\theta_\rmb$, and $\theta$ their angle variables. 
$\omega_\rmc$ and $\omega_\rmb$ are the angular velocities of cyclotron and bounce motion, and $\phi$ is the electric potential. 
The Hamilton's equation gives the following equation of motion on the magnetic coordinate.
%-------------------------------------------------------
\begin{eqnarray}
	\begin{array}{l}
		\ds \dd{\ell}{t} = \f{P_{||}}{m} \\
		\\
		\ds \dd{P_{||}}{t} = -\f{\mu q}{m}\pp{B}{\ell} - q\pp{\phi}{\ell} \\
		\\
		\ds \dd{\theta}{t} = \f{\mu}{m}\pp{B}{\psi} + \pp{\phi}{\psi} \\
		\\
		\ds \dd{\psi}{t} = -\pp{\phi}{\theta} \\
	\end{array},
\label{e:e.o.m.}
\end{eqnarray}
%-------------------------------------------------------
where the kinetic energy of drift motion has been omitted by the approximation $P_\theta = q\psi$. 
Below in order to observe particles diffusing with respect to $\psi$, the white noise perturbation of the azimuthal electric field $E_\theta \propto \p\phi/\p\theta$ is applied.

We note that (\ref{e:e.o.m.}) is simplified from a guiding-center equations derived in Ref.~\onlinecite{sato2015self} which include a geometric effect caused by non-orthogonality of a magnetic coordinate.
When the geometric effect is considered, the perturbation in azimuthal electric field ($\p\phi/\p\theta$) affects the parallel dynamics.
In this study, we omit the geometric effect for the sake of simplicity, hence the parallel dynamics is decoupled from the azimuthal electric field.

%-------------------------------------------------------
% CONSTRAINT AND FOLIATION BY MAGNETIC MOMENT
%-------------------------------------------------------
% \section{CONSTRAINT AND FOLIATION BY MAGNETIC MOMENT}%
We start by comparing orbits of particles foliated by magnetic moment with those of non-foliated particles that keep track of cyclotron orbits. 
The former is obtained by numerical solution of (\ref{e:e.o.m.}) and the latter by that of Newton's equation of motion.
First, in order to compare foliated and non-foliated particle orbits, we study the constancy of magnetic moment using a non-foliated particle. 
Next, we confirm the relationship between the constancy of magnetic moment and the macroscopic motion. 
Lastly, we compare foliated and non-foliated particle orbits on a variety of energy levels. 
All particle simulation parameters consider electron motion under the magnetic field of the RT-1 device~\cite{yoshida2006first}.
A particle is initially located on a magnetic field line that passes $r = 1$\,m and $z = 0$ where the magnetic field strength is $\sim 5\times10^{-3}$\,T.
The initial kinetic energy of the particle is $E_{\perp0} = E_{||0} = 5$\,eV, where $E_\perp$ and $E_{||}$ are parallel and perpendicular energy respectively, and subscript 0 denotes initial condition. 

\begin{figure}[htpb]
	\begin{center}
		\includegraphics*[width=0.5\textwidth]{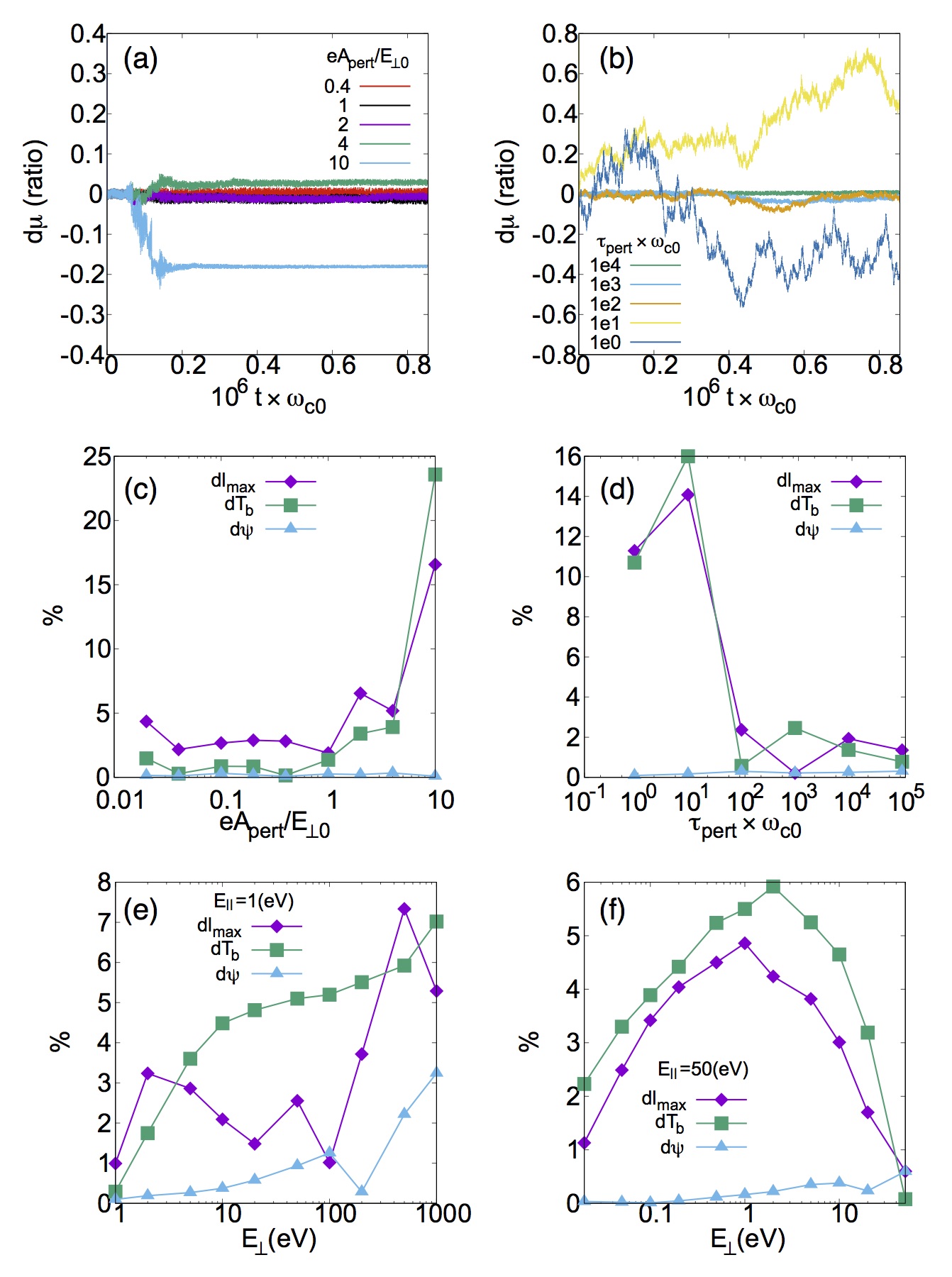}
	\end{center}
	\caption{
	(a)(b) The time evolution of magnetic moment $\mu(t)$ in relative difference rate $\rmd\mu\, (:= (\mu(t) - \mu(t=0))/\mu(t=0))$ from its initial value, under perturbation with various (a) amplitude (time scale is fixed $\tau_\mr{pert}\omega_{\rmc0} \sim 10^4$) and (b) time scale (amplitude is fixed $eA_\mr{pert}/E_{\perp0} = 1$). 
		(c)(d) The relative difference rate of bounce amplitude d$\ell_\mr{max}$(\%) (:= $|\ell_\mr{max\mu} - \ell_\mr{max}|/\ell_\mr{max}$), bounce period d$T_\mr{b}$(\%) (:= $|T_\mr{b\mu} - T_\mr{b}|/T_\mr{b}$), and magnetic flux function d$\psi$(\%) (:= $|\psi_{\mu} - \psi|/\psi$) between foliated (subscript $\mu$) and non-foliated (no subscript) particles, 
		under perturbation with various (c) amplitude and (d) time scale. (e)(f) d$\ell_\mr{max}$(\%), d$T_\rmb$(\%), d$\psi$(\%) with varied energy (e) ($E_{||} = 1\,\mr{eV}$, $E_{\perp} = 1\sim1000\,\mr{eV}$) and (f) ($E_{||} = 50\,\mr{eV}$, $E_{\perp} = 0.02\sim50\,\mr{eV}$).
	}
	\label{f:fig1}
\end{figure}

Figure~\ref{f:fig1}(a) shows the time evolution of the magnetic moment of a non-foliated particle as a difference rate from the initial value, with electrostatic potential perturbations of varying amplitude ($A_\mr{pert}$).
The figure shows that the magnetic moment differs from its initial value by less than 10\% with a weak perturbation ($eA_\mr{pert}/E_{\perp0} \le 4$). 
By a moderately strong perturbation of $eA_\mr{pert}/E_{\perp0} = 10$, the magnetic moment shifts to a $-20$\% value and gets fixed. 
This is due to the particle moving to a stronger magnetic field after a surge of perturbation disrupts its conservation of magnetic moment, allowing a stronger conservation of the new value. 
Such an occurrence may also be seen in $eA_\mr{pert}/E_{\perp0} = 4$. 
Perturbations with an amplitude of $eA_\mr{pert}/E_{\perp0} = 20$ and greater do not allow the particle to finish a gyrating orbit, and throws the particle outside the trapping zone of the magnetic field.

Figure~\ref{f:fig1}(b) shows the time evolution of the magnetic moment of a non-foliated particle with varying time scales of perturbation ($\tau_\mr{pert}$). 
The figure shows that the magnetic moment differs from its initial value by less than 10\% with a low frequency perturbation ($\tau_\mr{pert}\omega_{\rmc0} > 10$) where $\omega_{\rmc0} \sim 1.16\times10^{-9}\,\mr{s^{-1}}$ is the cyclotron frequency at $r = 1$\,m and $z = 0$. 
The difference of the magnetic moment increases largely with a high frequency perturbation $\tau_\mr{pert}\omega_{\rmc0} \le 10$.

Figure~\ref{f:fig1}(c) and (d) show the difference of foliated orbits to non-foliated orbits for values of the bounce amplitude $\ell_\mr{max}$, bounce period $T_\rmb$, and $\psi$. 
% The perturbation frequency and amplitude of the horizontal axis is correspondent to the perturbations in Fig.~\ref{f:fig1}(c) and (d). 
The difference largely increases with a high frequency perturbation of the same order as the Larmor period or with a moderately strong perturbation that allows particles to escape from the magnetic trapping zone.
The disruption of magnetic moment constancy disturbs the macroscopic structure given by the foliation, resulting in a higher difference between the foliated and non-foliated orbits.

Figure~\ref{f:fig1}(e) and (f) show the difference of foliated orbits to non-foliated orbits without perturbation under various initial kinetic energy. 
The combinations of $E_{||}$ and $E_\perp$ correspond to (e) varied kinetic energy and Larmor radius ($E_{||} = 1\,\mr{eV}$, $E_{\perp} = 1\sim1000\,\mr{eV}$) and (f) varied Larmor radius with a nearly constant kinetic energy ($E_{||} = 50\,\mr{eV}$, $E_{\perp} = 0.02\sim50\,\mr{eV}$). 
The figures show the effect upon the macro hierarchy caused by the enlargement of Larmor radius and the interaction of different periodic motions. 
In Fig.~\ref{f:fig1}(e), a larger Larmor radius by an increase in $E_\perp$ causes a varying magnetic field over the course of gyration, resulting in a higher difference. 
In Fig.~\ref{f:fig1}(f), the interaction of cyclotron motion and bounce motion is initially small with a low $E_\perp$, enlarges as $E_\perp$ increases, and returns small as $E_\perp$ nears $E_{||}$, resulting in a hill shape graph.
%-------------------------------------------------------
%-------------------------------------------------------

\begin{figure*}[htpb]
	\begin{center}
		\includegraphics*[width=1.0\textwidth]{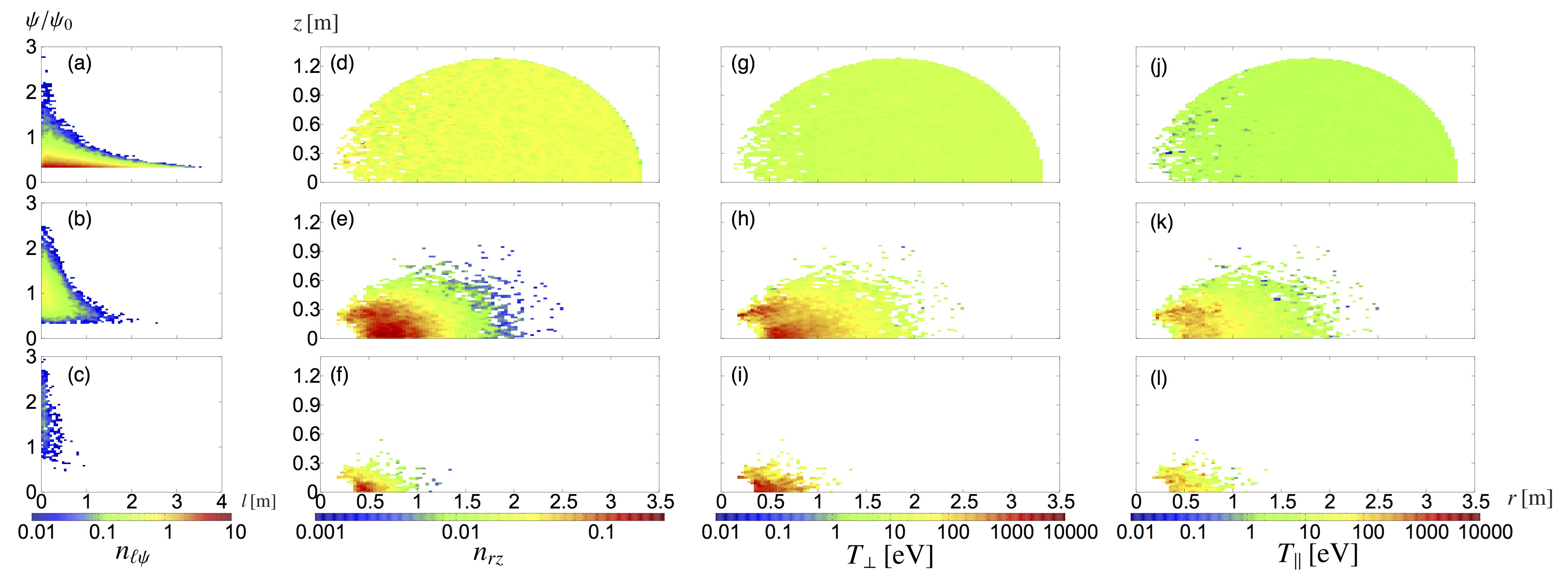}
	\end{center}
	\caption{
		The time evolution ($t=0$\,s (top), 0.02\,s (middle), 0.16\,s (bottom)) of density distribution (a)--(c) in the proper frame ($n_{\ell\psi}$), (d)--(f) in the laboratory frame ($n_{rz}$), (g)--(i) the  perpendicular temperature ($T_\perp = m\langle v_\perp^2\rangle/2$), and (j)--(l) the parallel temperature ($T_{||} = m\langle v_{||}^2\rangle/2$). 
	}
	\label{f:fig2}
\end{figure*}

\begin{figure*}[htpb]
	\begin{center}
		\includegraphics*[width=1.0\textwidth]{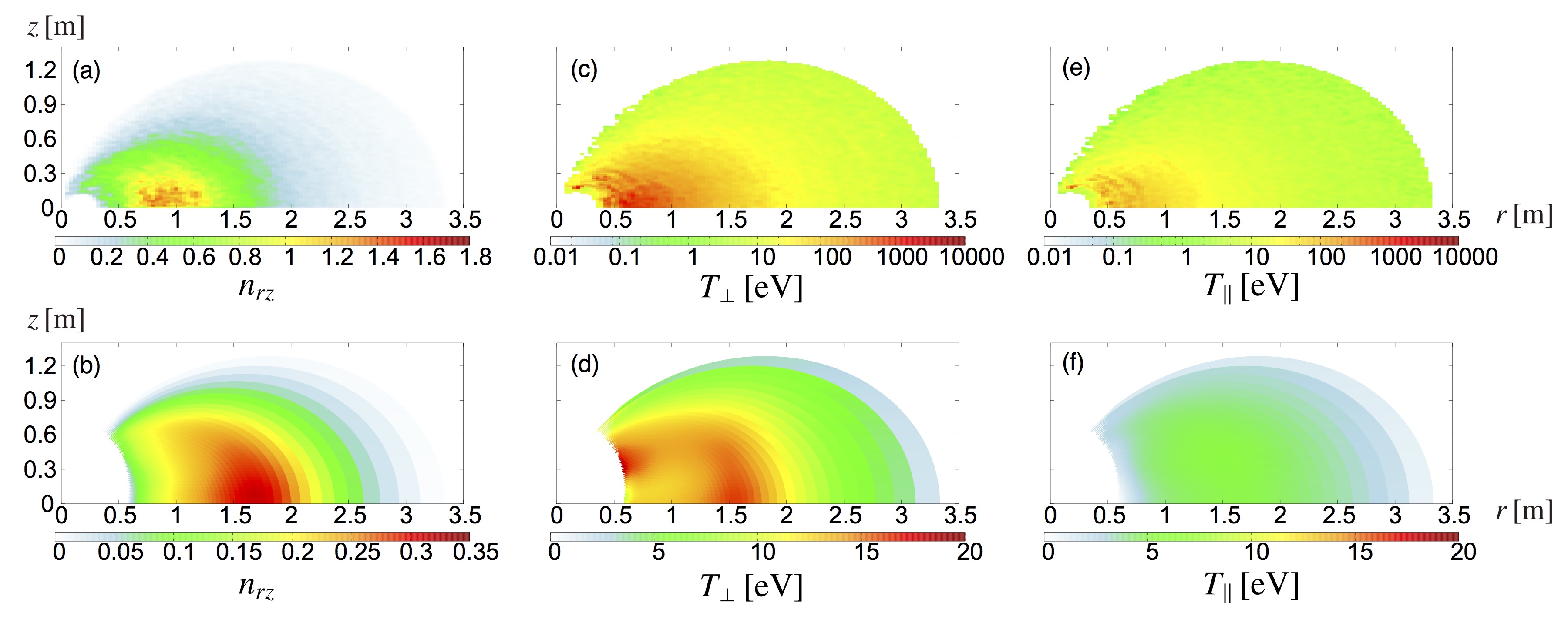}
	\end{center}
	\caption{
		The comparison of (a)--(b) $n_{rz}$, (c)--(d) $T_\perp$, and (e)--(f) $T_{||}$ between particle simulation (top) and diffusion model simulation (bottom). 
		The snapshots are taken at $t = 0.8$\,ms for the particle simulation and $t = 0.1$\,ms for the diffusion model simulation.
		The difference in the speed of evolution is due to the modeling of the diffusion coefficient from perturbation setting. 
	}
	\label{f:fig3}
\end{figure*}

%-------------------------------------------------------
% NUMERICAL SIMULATION OF FOLIATED PARTICLE MOTION
%-------------------------------------------------------
% \section{NUMERICAL SIMULATION OF FOLIATED PARTICLE MOTION}%
Next we present the results of many particle simulation calculated by (\ref{e:e.o.m.}). 
The computational cost of (\ref{e:e.o.m.}) is significantly reduced from that of Newton's equation of motion.
We observe the time evolution of particle distribution starting from an initial condition of a 10\,eV isotropic Maxwell-Boltzmann distribution.
Taking into account the above result, the perturbation amplitude and time scale are chosen so that particle orbits of (\ref{e:e.o.m.}) do not deviate from those of Newton's equation of motion:
$eA_\mr{pert}/E_{\perp0} = 0.125$ and $\tau_\mr{pert}\omega_{\rmc0} = 1\times10^{5}$ for Fig.~\ref{f:fig2}, and 
$eA_\mr{pert}/E_{\perp0} = 0.4$ and $\tau_\mr{pert}\omega_{\rmc0} = 2\times10^{5}$ for Fig.~\ref{f:fig3}. 
The particle number of the simulations is 70903 for Fig.~\ref{f:fig2} and 46016 for Fig.~\ref{f:fig3}.

Figure~\ref{f:fig2}(a)--(f) show the time evolution of the density distribution in the proper frame ($\ell-\psi$ coordinate) and in the laboratory frame ($r-z$ coordinate). 
Here, $\psi$ is normalized by $\psi_0 := \psi(r=1\,\rmm,\, z=0)$.
Diffusion on the proper frame is observed as inward diffusion in the laboratory frame. 
Particles start from an inhomogeneous distribution in the proper frame, which corresponds to the homogeneous distribution in the laboratory frame. 
% The density gradient disperses with respect to $\psi$ by perturbation, 
% and the particle number per unit flux is equalized in the final state, which is consistent with the preceding theoretical prediction~\cite{Hasegawa2005,kesner2011fluctuation}.
% Since $\psi$ increases moving inwards, this flattening corresponds to the inward diffusion, as explicitly observed in $r-z$ coordinate (Fig.~\ref{f:fig2}(d)--(f)).
  Diffusion in $\psi$ diminishes the density gradient in the proper coordinates and the particle number per unit flux tube is progressively homogenized. 
  Since the flux tube volume becomes thinner as one moves toward the center of the dipole, the flattening in the proper coordinates is seen as a steepening density profile in the laboratory frame (Fig.~\ref{f:fig2}(d)--(f)). 
  This scenario is consistent with theoretical predictions~\cite{Hasegawa2005,kesner2011fluctuation}.  
Particle distribution with respect to $\ell$ is squeezed into the equatorial plane ($\ell=0$) by the mirror effect. 
% Such a time evolution seen in the laboratory frame is observed as inward diffusion.

Figure~\ref{f:fig2}(g)--(l) show the time evolution of the parallel and perpendicular temperature ($T_{||} = m\langle v_{||}^2\rangle/2$ and $T_\perp = m\langle v_\perp^2\rangle/2$) with respect to the magnetic field lines. 
The figure shows that particles are heated anisotropically. 
Initially, the distributions of both the parallel and perpendicular temperature are homogeneous, and their values are of the same order. 
Then both of the temperature distributions evolve to form a peak area by perturbation. 
The peak area corresponds to the high-$\psi$ edge of the density peak.  
This is due to the acceleration of particles as they move in the positive $\psi$ direction, and a higher statistical chance of high energy particles with higher particle number. 
While particles are accelerated in the perpendicular direction as they diffuse inwards, they may get accelerated or decelerate in the parallel direction depending on which point of the bounce orbit they are at.
Such a difference in the acceleration mechanism causes the peak of the parallel and perpendicular temperature to have a different position and value. 
Thus the anisotropic heating of inward diffusion is confirmed by particle delineation.
%-------------------------------------------------------
%-------------------------------------------------------

Next we compare the particle simulation results with simulation results gained by the diffusion model by Sato~\cite{Sato2015}. 
The aforementioned perturbation parameter setting corresponds to the same order of diffusion coefficient used in the diffusion model simulation.
The assumed perturbation waveforms are, however, different between the particle simulation and the diffusion model simulation;
the former assumes a rectangular waveform and the latter assumes a triangular waveform~\cite{Sato2015}. 
Figure~\ref{f:fig3}(a) and (b) show the comparison of the laboratory frame density between the particle simulation and the diffusion model simulation. 
In both models, diffusion with respect to $\psi$ is observed as inward diffusion in the laboratory frame with particle distribution getting squeezed into the equatorial plane as the diffusion progresses. 
    The difference of the time steps between the figures ($t=0.8$\,ms for the particle simulation and $t=0.1$\,ms for the diffusion model simulation) is due to the different values of the diffusion coefficient. 
  In particular, by comparing the strength of diffusion in the two simulations, we find that the distribution calculated with the diffusion model at $t=0.1$\,ms roughly corresponds to that obtained by particle simulation at $t=0.8$\,ms.
    % overestimate of the diffusion coefficient in the diffusion model
    % (value of a diffusion coefficient estimated by perturbation parameters may vary by several factors via a stochastic modeling procedure).
    % % Although the perturbation parameters in the particle simulation was determined so that the resulting diffusion coefficient corresponds to that for the diffusion model simulation, stochastic modeling from the perturbation allows several times smaller (or larger) diffusion coefficient.
    % % Thus, the distributions by the diffusion model at $t=0.1$\,ms correspond to the particle simulation at $t=0.8$\,ms.
    % % Stochastic modeling from the perturbation results in a diffusion coefficient that is several times smaller (or larger).
    % Thus, the distributions by the diffusion model at $t=0.1$\,ms correspond to the particle simulation at $t=0.8$\,ms.
    As shown in the previous study on the diffusion model~\cite{Sato2015}, the density distribution will be further squeezed into the equatorial plane as time advances (i.e., $t>0.1$\,ms), creating vertically a thin structure. 
    Therefore, we expect that such a change will observed in the particle simulation at $t>0.8$\,ms as well.
    However, particle simulations longer than $0.8$\,ms were not conducted in this study.
% The difference of the timestamps between the figures ($t=0.8$\,ms for the particle simulation and $t=0.1$\,ms for the diffusion model simulation) is due to the overestimate of the diffusion coefficient in the diffusion model
% (value of a diffusion coefficient estimated by perturbation parameters may vary by several factors via a stochastic modeling procedure).
% % Although the perturbation parameters in the particle simulation was determined so that the resulting diffusion coefficient corresponds to that for the diffusion model simulation, stochastic modeling from the perturbation allows several times smaller (or larger) diffusion coefficient.
% % Thus, the distributions by the diffusion model at $t=0.1$\,ms correspond to the particle simulation at $t=0.8$\,ms.
% % Stochastic modeling from the perturbation results in a diffusion coefficient that is several times smaller (or larger).
% Thus, the distributions by the diffusion model at $t=0.1$\,ms correspond to the particle simulation at $t=0.8$\,ms.
% As shown in the previous study on the diffusion model~\cite{Sato2015}, the density distribution will be further squeezed into the equatorial plane as time advances (i.e., $t>0.1$\,ms), creating vertically thin structure. 
% Therefore, we expect that such a change will observed in the particle simulation at $t>0.8$\,ms as well.
% However, the particle simulation longer than $0.8$\,ms was not conducted in this study.
Figure~\ref{f:fig3} (c)--(f) show the comparison of temperature distribution between the particle model and diffusion model. 
Both simulations show good agreement on the anisotropic heating resulting in $T_\perp > T_{||}$, whereas the diffusion model gives the larger anisotropy. 

% Differences appear due to the geometric factor, friction factor, and perturbation along magnetic field lines in the diffusion model. 
% In both cases of temperature parallel and perpendicular with respect to the magnetic field lines, the diffusion model show a stronger squeeze in the equatorial plane compared to the particle model, due to the geometric factor and friction factor. 
% For the parallel temperature, the peak is not shaped on the equatorial plane in the diffusion model unlike in the particle model, but on an area biased towards a larger $\ell$, due to the perturbation along field lines in the diffusion model. 
% For the perpendicular temperature, the diffusion model and particle model share the double peak formation.
%-------------------------------------------------------
%-------------------------------------------------------

%-------------------------------------------------------
% CONCLUSION
%-------------------------------------------------------
% \section{CONCLUSION}%
% Particles in magnetospheres diffuse inwards, getting accelerated in the process, and self-organizes into an inhomogeneous structure. 
% Such phenomenon may be described as the diffusion on a foliated phase space (\textit{i.e.} proper frame). 
% In the previous research, the picture of particle motion as a diffusion on the proper frame with foliated phase space has been given the Boltzmann distribution on the foliation of adiabatic invariants, and a diffusion model with geometric factors.
% In the present work, an additional particle delineation was given through particle simulation in order to unveil the marginal scale of particle motion connecting micro-scale motions to macro-scale self-organization. 
% In the present work, an additional particle delineation was given through particle simulation in order to unveil the marginal scale of particle motion connecting micro-scale motions to macro-scale self-organization. 
In conclusion, the inward diffusion process was investigated by numerical simulation of particles on the foliated phase space.
Clumping and anisotropic heating of the particles were observed. 
The obtained density and temperature profiles showed good agreement with the diffusion model in the previous study.
% The process and effect of the structurization by the foliation of magnetic moment was investigated. 
% The anisotropic heating process was confirmed by particle delineation.
% The particle and diffusion model simulations showed a good agreement.
Such structurization and anisotropic heating are also consistent with the laboratory experiment results~\cite{yoshida2010magnetospheric,boxer2010turbulent,saitoh2011high,Kawazura2015}.

% \section*{acknowledgments}
This work was supported by JSPS KAKENHI Grant No. 23224014. 

\setcitestyle{numbers,square}
\bibliographystyle{apsrev4-1}
\bibliography{references}

\end{document}